\documentclass[aps,pra,twocolumn,groupedaddress]{revtex4}
\usepackage{graphicx}
\usepackage{amssymb}
\usepackage{subfigure}
\usepackage{amsmath}
\usepackage{amsfonts}
\usepackage{bm}
\usepackage{color}
\graphicspath{%
    {converted_graphics/}
    {C:/Users/PH/Desktop/UFSC/Dados/}
    {C:/Users/PH/Desktop/UFSC/Deutsch/}
    {C:/Users/PH/Desktop/}
    {C:/Users/PH/Desktop/UFSC/Deutsch/Dados-fim/}
    {C:/Users/PH/Desktop/UFSC/Deutsch/Natal/}
}
\begin{document}


\title{Deterministic Quantum Computation With One Photonic Qubit}

\author{M. Hor-Meyll}
\affiliation{Instituto de F\'{\i}sica, Universidade Federal do Rio de
Janeiro, Caixa Postal 68528, Rio de Janeiro, RJ 21941-972, Brazil}
\author{D. S. Tasca}
\affiliation{Instituto de F\'{\i}sica, Universidade Federal do Rio de
Janeiro, Caixa Postal 68528, Rio de Janeiro, RJ 21941-972, Brazil}
\author{S. P. Walborn}
\affiliation{Instituto de F\'{\i}sica, Universidade Federal do Rio de
Janeiro, Caixa Postal 68528, Rio de Janeiro, RJ 21941-972, Brazil}
\author{M. M. Santos}
\affiliation{Instituto de F\'{\i}sica, Universidade Federal de Uberl\^andia, Caixa Postal 593, 38400-902, Uberl\^andia, MG, Brazil}
\affiliation{Instituto de Ci\^{e}ncia, Engenharia e Tecnologia,Universidade Federal dos Vales do Jequitinhonha e Mucuri, Rua do Cruzeiro 01, Jardim S\~ao Paulo, 39803-371, Teofilo Otoni, Minas Gerais, Brazil}
\author{E. I. Duzzioni}
\affiliation{Instituto de F\'{\i}sica, Universidade Federal de Uberl\^andia, Caixa Postal 593, 38400-902, Uberl\^andia, MG, Brazil}
\affiliation{Departamento de Física, Universidade Federal de Santa Catarina, Caixa Postal 476, 88040-900, Florianopolis, SC Brazil}
\author{P. H. Souto Ribeiro}
\email{phsr@if.ufrj.br}
\affiliation{Instituto de F\'{\i}sica, Universidade Federal do Rio de
Janeiro, Caixa Postal 68528, Rio de Janeiro, RJ 21941-972, Brazil}
\affiliation{Departamento de Física, Universidade Federal de Santa Catarina, Caixa Postal 476, 88040-900, Florianopolis, SC Brazil}

\begin{abstract}
We show that deterministic quantum computing with one qubit (DQC1) can be experimentally implemented
with a spatial light modulator, using the polarization and the transverse spatial degrees of freedom of light. The scheme allows the computation of the trace of a high dimension matrix,
being limited by the resolution of the modulator panel, and the technical imperfections. 
In order to illustrate the method, we compute the normalized trace of unitary matrices, and implement the Deutsch-Jozsa algorithm. The largest matrix that can be manipulated with our set-up is 1080$\times$1920, which is able to represent a system with approximately 21 qubits.  
\end{abstract}



\maketitle
\section{Introduction}

The field of quantum information has been strongly motivated by the demonstration that some quantum algorithms have an improved performance in comparison to their classical versions. More recently, the deterministic quantum computation with one quantum bit (DQC1) model was introduced with the aim of exploring the computational speed-up in high temperature ensemble quantum computation \cite{1}. Although this model of computation is not universal, it enables quantum speed up to solve certain problems, such as the Shor factorization algorithm \cite{2}, the measurement of the average fidelity decay of a quantum map \cite{3}, the trace calculus of an arbitrary unitary evolution \cite{4}, and the approximation of the Jones Polinomial \cite{5}. The importance of the DQC1 model of computing is that it requires little or no entanglement between the qubits of the system \cite{5a} to evaluate the trace of a unitary operator, an operation that is not efficiently implemented by a classical computer \cite{4}. The experimental implementation of this model of computation has already been made in optical systems \cite{6} and Nuclear Magnetic Resonance \cite{7,8}.  

We present an experimental scheme for the implementation of the DQC1 protocol in an optical scenario. In this paradigmatic model, the normalized trace of a unitary matrix is computed by using the computational power of only one qubit in a pure state, and a collection of qubits in a completely mixed state. The information about the normalized trace of the unitary matrix is transferred to the qubit state through conditional operations. For this purpose we use a phase-only Spatial Light Modulator (SLM), which performs polarization-controlled position-dependent phase shifts. Proper polarization measurements return the result of the computation. A similar approach was recently used to demonstrate the use of a SLM and polarized light to estimate integrals \cite{9}. In addition to several possible applications of the SLM \cite{10} and its use as a quantum channel acting on the polarization \cite{105}, here we are interested in the implementation of quantum algorithms. Some quantum algorithms have already been implemented or simulated using the SLM, such as the Deutsch algorithm\cite{10a}, Deutsch-Jozsa algorithm \cite{11,12}, the Grover algorithm \cite{12}, and the quantum walk \cite{13}.  Moreover, there has been an implementation of the Deutsch algorithm for two qubits encoded in an optical system using an operation controlled by polarization \cite{13a}, but there an interferometer was needed to control the spatial mode.

In order to illustrate the implementation of our method, some examples of trace calculation are presented
for a few types of matrices and also the realization of the Deutsch-Jozsa algorithm. The oracle function is prepared in the SLM as a matrix of adjustable dimension, up to its resolution limits. For instance, for our High-Definition (HD) panel and using full resolution of  1080 x 1920 pixels, we could implement an oracle function with approximately 2$^{21}$ inputs, which would correspond to an input composed of 21 qubits.
We show that the performance of the method in the present realization is limited by
polarization dephasing effects \cite{105}, and other sources of noise like fluctuations in the phase modulation and photon number statistics. 

\section{DQC1 using a photonic qubit}

\begin{figure}[tbp] 
\centerline{\includegraphics[width=2.5in,height=2.23in,keepaspectratio]{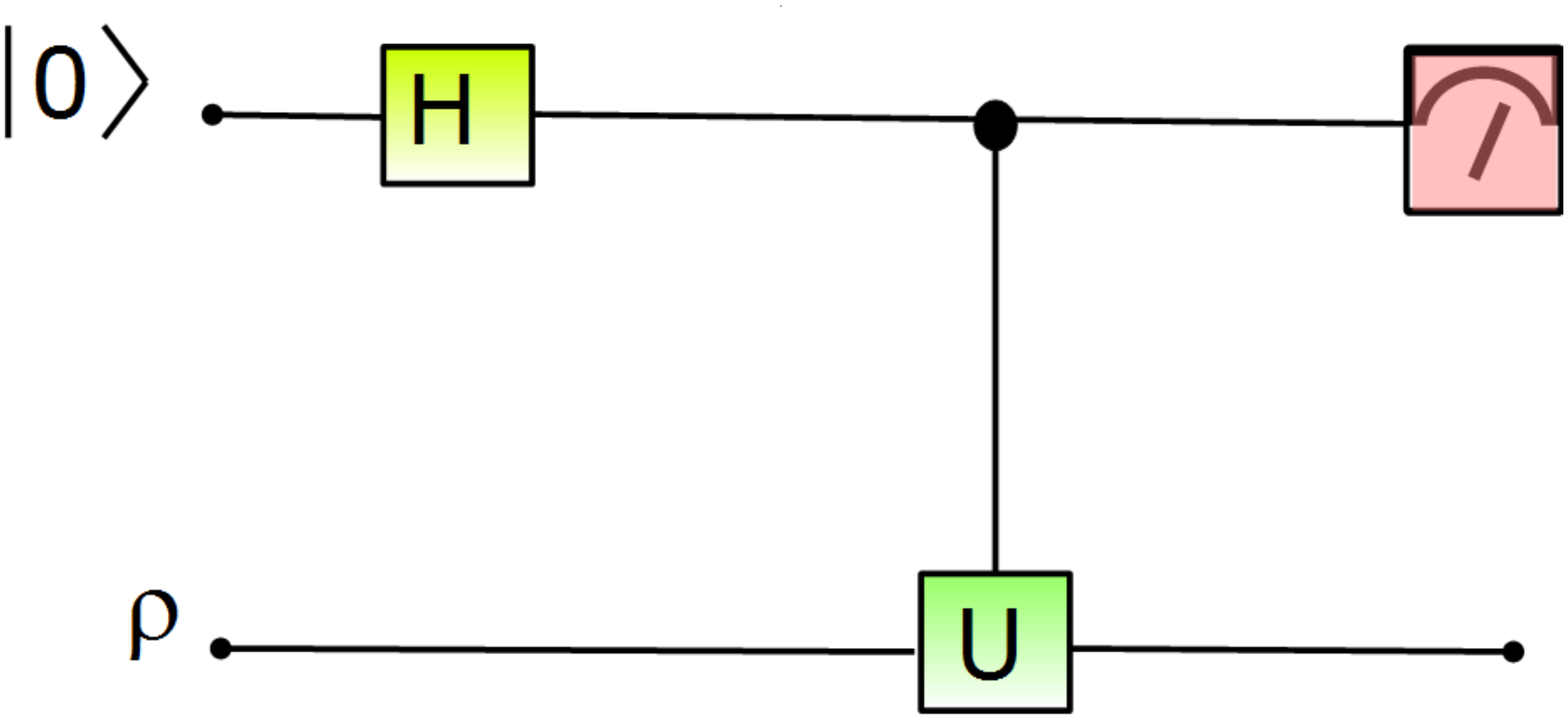}}
\caption{DQC1 circuit. The action of the Hadammard gate on the state $|0\rangle$ gives the 
state $|+\rangle = (|0\rangle + |1\rangle)/\sqrt{2}$, while the controlled unitary transformation U acts on the state $\rho$.}
 \label{fig:FIG1}
\end{figure}

We start the discussion by explaining how we can use the SLM to implement 
the DQC1 circuit as sketched in Fig. \ref{fig:FIG1}.
We use a light beam with a given transverse wavefront profile as the system $\rho = \rho_t$, 
and its polarization state as the control qubit. The initial photonic state can be described by:

\begin{equation}
\rho_{i}= |+\rangle \langle +| \otimes \rho_{t} ,
\label{input}
\end{equation}
where the polarization state 
$|+\rangle = (|0\rangle + |1\rangle$)/$\sqrt{2} \equiv (|H\rangle + |V\rangle$)/$\sqrt{2}$, with $|H\rangle$
($|V\rangle$) representing the linear horizontal (vertical) polarization state and the 
state of the transverse wavefront is described by $\rho_{t}$.\\ 

Both the polarization and the transverse wavefront are properties of the same light beam.
It is convenient to describe the wavefront in terms of a discrete basis:

\begin{equation}
\label{rhobase}
\rho_{t} = C \sum_{i,j} |x_i, y_j \rangle \langle x_i, y_j|,
\end{equation}
where each state of the basis describes the wavefront in a surface given
by the pixel area of a SLM, and $C$ is a normalization constant.
The SLM will be used to implement the controlled operation. 
$x_i$ and $y_j$ are the coordinates of pixel $i,j$. 
Notice that in this description the state of the wavefront is maximally mixed. 
We will discuss later how this state can be obtained experimentally.

Let us define an operator describing the action of the SLM on states
of the form given by Eq. (\ref{input}) as \cite{1305}:

\begin{equation}
S = |H\rangle \langle H| \otimes U + |V\rangle \langle V| \otimes {1\!\!1},
\label{slmop}
\end{equation}
where 
\begin{equation}
U = \sum_{i,j} \mbox{e}^{-i \phi(x_i,y_j)}|x_i,y_j\rangle\langle x_i,y_j|,
\label{U}
\end{equation}
and $\phi(x_i,y_j)$ is a real function.

The SLM used in our experiment only modulates the horizontal 
polarization component of the input light beam. The function $\phi(x_i,y_j)$ is programmed to apply a phase between $0$ 
and $2\pi$ in each SLM pixel located at $(x_i,y_j)$.

We use Eqs. (\ref{input}-\ref{slmop}) to obtain the photonic
state after incidence on the SLM:

\begin{eqnarray}
\label{rhof}
\rho_{f} &=& S \rho_i S^{\dagger} = S [|+\rangle \langle +| \otimes \rho_{t}] S^{\dagger} \\ \nonumber
&=& (1/2) (|H\rangle \langle H| U \rho_t U^\dagger+
|V\rangle \langle V|\rho_t + \\ \nonumber
&+& |H\rangle \langle V| U \rho_t {1\!\!1} + |V\rangle \langle H|{1\!\!1} \rho_t U^\dagger),
\end{eqnarray}
and using the decompositions in Eqs. (\ref{rhobase}) and (\ref{U}), we calculate the partial trace over the
spatial degrees of freedom:

\begin{eqnarray}
\label{traco}
\rho_{pol} \equiv \mbox{Tr}_{t}[\rho_f] &=& \frac{1}{2} 
\bigg( |H\rangle \langle H|+|V\rangle \langle V|+ \\ \nonumber
&+&|H\rangle \langle V|C\sum_{i,j} \mbox{e}^{-i \phi(x_i,y_j)}  + \\ \nonumber
&+& |V\rangle \langle H| C\sum_{i,j} \mbox{e}^{+i \phi(x_i,y_j)}\bigg).
\end{eqnarray}

In terms of matrix representation the state can be written in the basis $\{|H\rangle,|V\rangle\}$ as:

\begin{equation} 
\rho_{pol} = \frac{1}{2}\left(\begin{array}{ccc}
1 & C\sum_{i,j} \mbox{e}^{-i \phi(x_i,y_j)}  \\
C\sum_{i,j} \mbox{e}^{+i \phi(x_i,y_j)}  & 1 
\end{array} \right).
\label{rhopol}
\end{equation} 

This result shows that the information about the spatial modulation
is transferred to the coherences of the polarization state, in terms
of the average of the modulation distribution. Therefore, the expectation
value $\langle \sigma_x \rangle$ of the Pauli operator $\sigma_x$ for this
state gives:

\begin{equation}
\label{sigmax}
 \langle \sigma_x \rangle = 
C\sum_{i,j} \,\, \mbox{cos}[\phi(x_i,y_j)],
\end{equation}

which is the sum over the real parts of the modulation phases. We also have access to the sum over the imaginary parts of the modulation phases through the measurement of $\sigma_y$:

\begin{equation}
\label{sigmay}
\langle \sigma_y \rangle =
C \sum_{i,j} \,\, \mbox{sin} [\phi(x_i,y_j)].
\end{equation}

We can interpret the modulation phases $e^{-i \phi(x_i,y_j)}$ for each pixel as the diagonal elements of a matrix. Therefore, our scheme provides a method for calculating the 
normalized trace of this matrix through the measurement of  $\langle \sigma_x \rangle$ and $ \langle \sigma_y \rangle$.

\subsection{Implementation of the Deutsch-Jozsa algorithm}

\begin{figure}[tbp] 
 
 \includegraphics[bb=0 0 842 595,width=3in,height=2.92in,keepaspectratio]{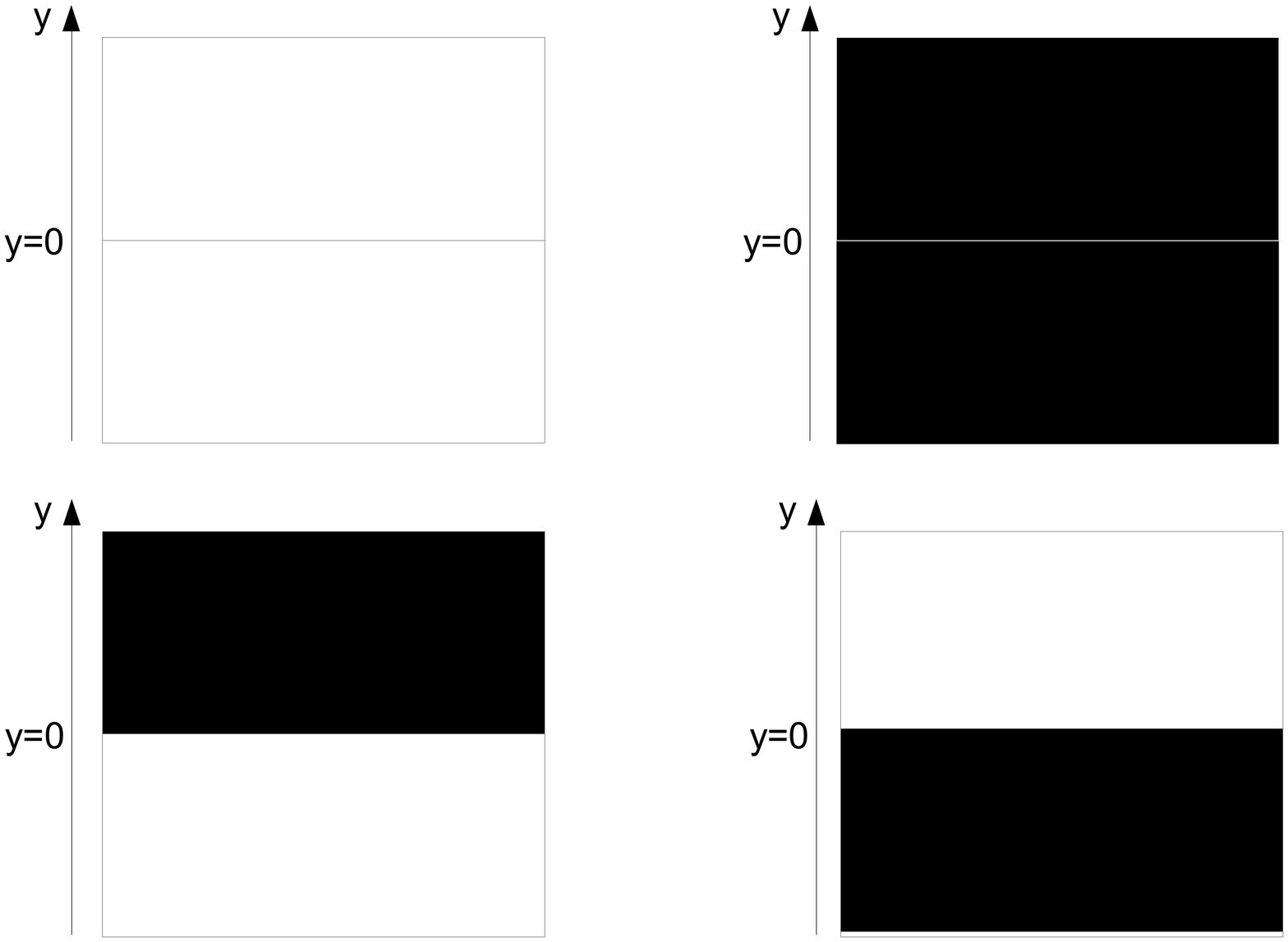}

  \caption{Representation of the possible modulations for $d=2$. White corresponds to 
a modulation phase $0$, and black corresponds to phase $\pi$.}
  \label{slm2}
\end{figure}

To illustrate further utility of the scheme, let us consider the implementation
of the Deutsch-Jozsa algorithm. In this paradigmatic quantum
algorithm, one wishes to test if an oracle function is constant or balanced \cite{DJ}.
The oracle function is implemented on the SLM. In the simplest case, also known as the Deutsch
algorithm \cite{D}, the SLM surface is divided in only two equal parts, where the upper part is given by $y_j \geq 0$ and the lower part is given by $y_j < 0$.
This corresponds to dimension $d=2$, as illustrated in Fig. \ref{slm2}.
For this algorithm, we use only two possible modulation phases for each half of the SLM: $0$ or $\pi$.
For $\phi(x_i,y_j) = 0$, $\langle \sigma_x \rangle = +1$, and for  $\phi(x_i,y_j) = \pi$, 
$\langle \sigma_x \rangle = -1$.
Therefore, for a constant function we have $\langle \sigma_x \rangle = \pm 1$
and for a balanced function we have $\langle \sigma_x \rangle = 0$.

The algorithm is easily generalized (Deutsch-Jozsa algorithm) to any desired number of divisions of the SLM
surface, up to the limit of a single pixel per division. In this case, we can have
1080 $\times$ 1920 $=$ 2073600 cells that could be programmed individually with $0$ or $\pi$.
This is equivalent to testing a function with an input of approximately 21 qubits.
The expected result for the Deutsch-Jozsa algorithm is the same as for 2 divisions,
meaning that every kind of modulation map that modulates half of the surface with
$0$ and half with $\pi$ implements a balanced function and should return zero as result.

\section{Experiment}

\begin{figure}
\vspace*{1.5cm}
\leftline{\includegraphics[bb=87 195 738 436,width=3in,height=3.3in,keepaspectratio]{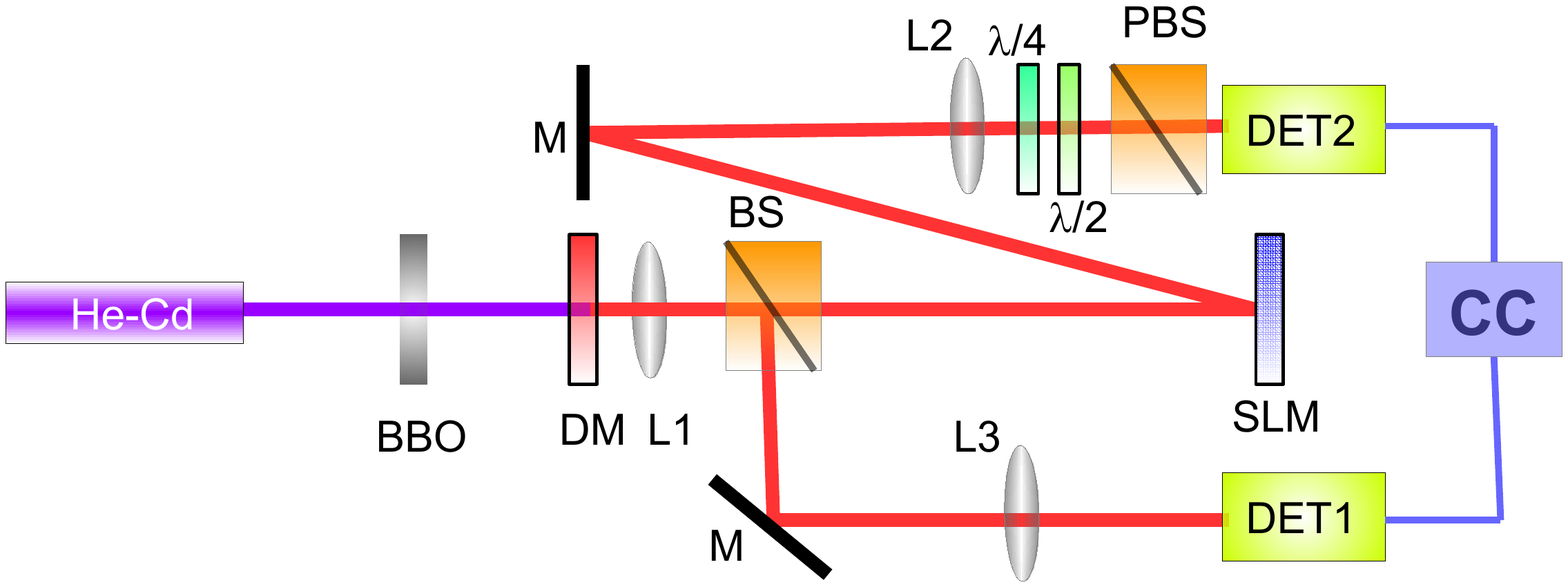}}
\vspace*{0.7cm}
  \caption{Experimental set-up. He-Cd is a Helium-Cadmium laser,
BBO is a nonlinear crystal, DM is a dichroic mirror, M represents mirrors, 
L1, L2 and L3 are lenses, $\lambda/4$ is a quarter wave plate, $\lambda/2$ is a half wave plate,
BS is a 50:50 beam splitter, PBS is a polarizing beam splitter, SLM is the spatial light modulator,
DET1 and DET2 are single photon counting modules, and CC is a coincidence detection circuit.}
  \label{fig:FIG2}
\end{figure}

The experimental set-up is sketched in Fig. \ref{fig:FIG2}.
A helium-cadmium (He-Cd) laser oscillating at 325nm is used to pump  a type I BBO nonlinear
crystal. It produces pairs of photons via parametric down-conversion, and we adjust the phase matching angle
to obtain collinear twin beams at the degenerate wavelength of 650nm. The down-converted signal and idler
beams are separated with a 50:50 beam splitter, and detected using 10nm bandwidth interference 
filters placed in front of the detectors. 
The idler beam is sent directly  to a single photon counter labeled DET1, and the signal beam 
is sent to a spatial light modulator (SLM) and polarization optics before detection by a single photon counter labeled DET2. Coincidence detection is used to post-select time correlated events signaling the arrival of a twin  photon pair. In this way, the idler photon at DET1 heralds the presence of the signal photon. 
 The signal photon propagates through L1, which is a lens implementing an optical Fourier transform 
mapping the far-field distribution in the crystal plane onto the SLM. 
Using this lens system we avoid much of the free propagation effects 
of the beams, minimizing unwanted diffraction and transverse spatial cross-correlations.
The idler beam also propagates through L1 before splitting in the beam splitter. Its far-field
distribution is mapped on an intermediate plane, just like the signal but without an SLM.
In our experiment, we use a reflective full HD 1080 $\times$ 1920 pixels, phase-only  SLM made by Holoeye 
Photonics. We are able to access each pixel individually and program any modulation phase
ranging from $0$ to $2\pi$. A specific function $\phi(x_i,y_j)$ characterizing the modulation distribution in
the plane of the SLM is programmed, and the signal field wavefront acquires this phase conditioned on the polarization. The input beam is prepared in a linearly polarized state
along the diagonal direction: $|+\rangle = (|H\rangle + |V\rangle$)/$\sqrt{2}$, so that the horizontal
component is modulated and the vertical is not. After interaction with the SLM, we measure the resulting polarization state given by Eq. (\ref{rhopol}), in two different basis, obtaining information about the 
modulation of the spatial profile.  Lens L2 forms the image of the SLM surface in the detection plane.
Lens L3 in the idler beam forms the image of the intermediate plane were the far field were mapped
previously by L1. In this way, the spatial propagations of signal and idler beams are equivalent,
except for the presence of the SLM in the signal.
The polarization analysis is made with a quarter wave plate $\lambda/4$,
followed by a half wave plate $\lambda/2$ and a polarizing beam splitter (PBS). 
We register both the single photon and two-photon coincidence counts. 

Due to the spatial correlations, depending on how the idler photon is detected, 
the heralded signal photon can be prepared to
have different spatial properties \cite{review}. For a small idler detection area,
the heralded signal beam becomes spatially coherent, which means that Eq. (\ref{rhobase}) is not suitable for
describing its transverse wavefront. For a large detection area  of the idler photon, the heralded signal beam 
is spatially incoherent, and is well described by Eq. (\ref{rhobase}). The use of twin photons
and coincidence detection thus allows the preparation of states with controllable purity.

\begin{figure}
\centerline{\includegraphics[bb=2 6 241 268,width=2.5in,height=4.49in,keepaspectratio]{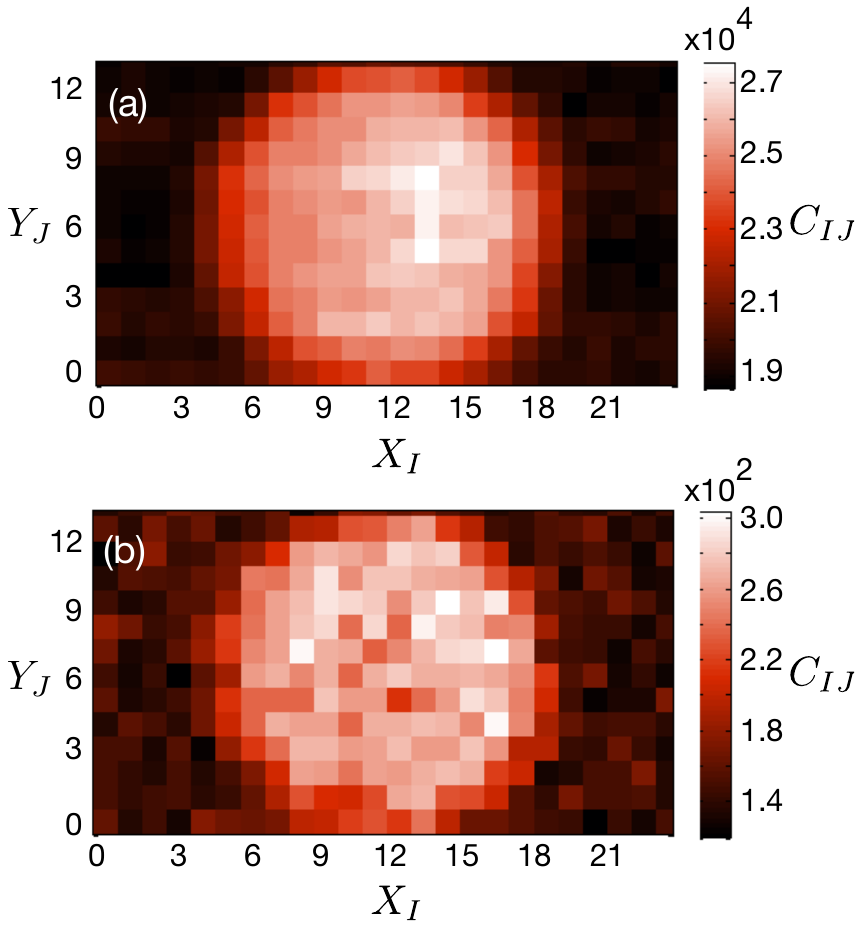}}
\caption{Transverse profile of the beam on the SLM. (a) Single photon counting distribution (counts/s). (b) Coincidence counting rate (coincidences/s). The transverse coordinates $X_I$ and $Y_J$ refer to the coordinates of detection cells containing an array of $(80 \times 80)$ pixels. $C_{IJ}$ is the single/coincidence counting rate in the cell located at $(X_I,Y_J)$}.
\label{beam}
\end{figure}

It is important to characterize the spatial intensity distribution of the light beam
interacting with the SLM panel, as it may have some influence on the computation. Fig. \ref{beam}
shows the intensity profile of the signal light beam used, already triggered by the idler. This profile will be taken into account in the calculation of the trace of a matrix using our scheme.

\section{Results}

\subsection{Implementation of the Deutsch-Jozsa Algorithm}

\begin{figure}[tbp] 
  \centering
 \centerline{\includegraphics[bb=39 42 510 814,width=2.5in,height=5in,keepaspectratio]{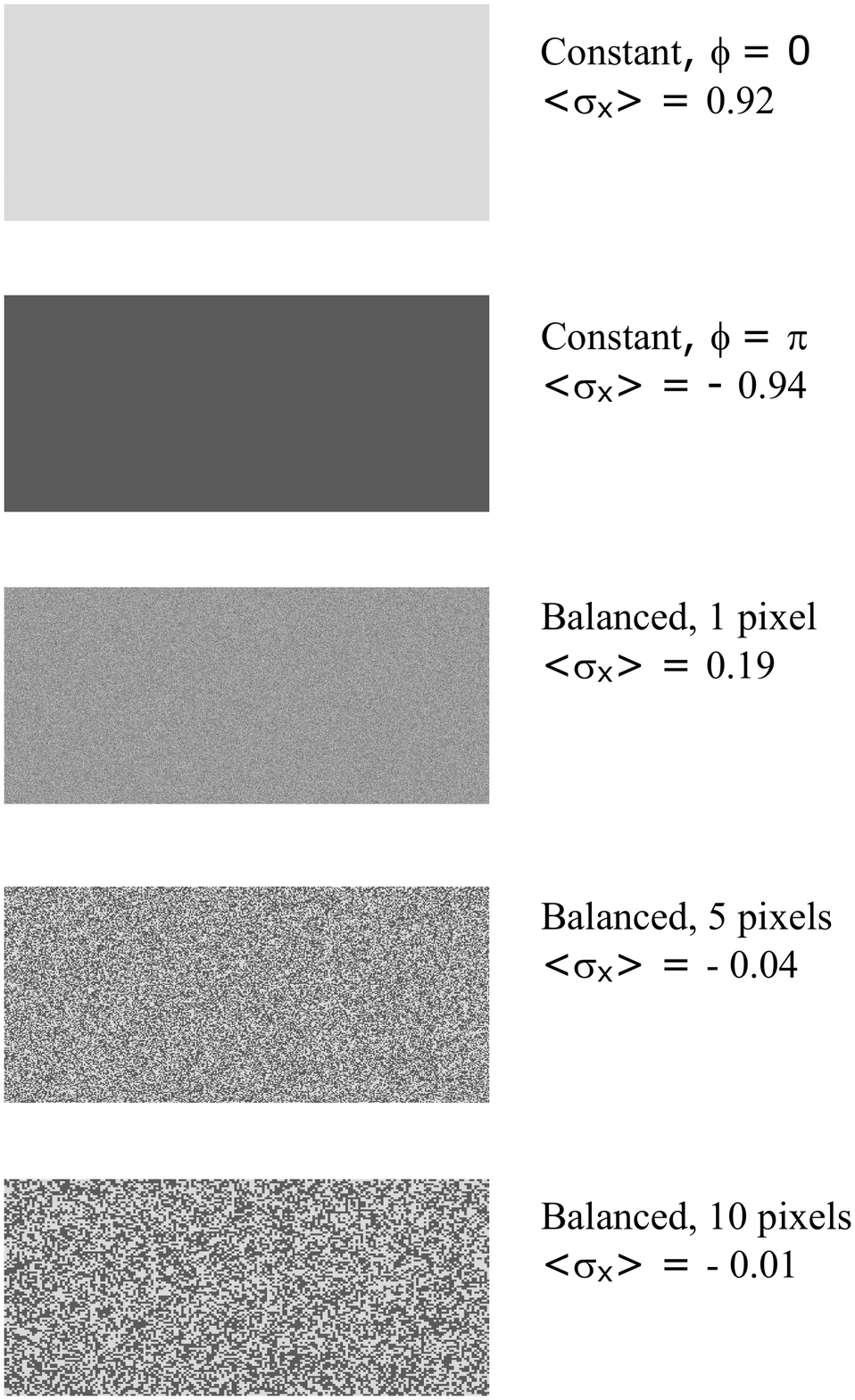}}

    \caption{Images displayed on the SLM panel for constant and balanced functions and the respective measured values of $\langle \sigma_x \rangle$. The balanced functions are implemented through (1 $\times$ 1), (5 $\times$ 5), and (10 $\times$ 10) square pixels cells randomly distributed on the panel.}
  \label{fig2}
\end{figure}

We begin by presenting the results obtained in the implementation of the Deutsch-Jozsa algorithm,
as it is a special (and simpler) case of the general calculation of the normalized trace of a matrix.
In Fig. \ref{fig2} we show the surface modulation of the SLM and the measured values of 
$\langle \sigma_x \rangle$. For the constant functions the ideal value should be $\langle \sigma_x \rangle = +1$
for $\phi = 0$ and $\langle \sigma_x \rangle = -1$ for $\phi = \pi$, and we observe a good agreement
with the measurements. For the balanced functions, we apply SLM masks with half of the cells modulated
with $\phi = 0$ and half with $\phi = \pi$ distributed randomly on the SLM surface. The randomness reduces the need of a precise knowledge of the spatial distribution of the light beam on the surface of the SLM. The results are shown in Fig. \ref{fig2} for square cells containing (1 $\times$ 1), (5 $\times$ 5), 
and (10 $\times$ 10) pixels.  
We can see that for all resolutions the measured value of $\langle \sigma_x \rangle$ is very close to
zero, showing that the algorithm works well even when using the full resolution of the SLM. For the purpose of
deciding if the oracle function is constant or balanced, a relatively high degree of uncertainty is
tolerated, since one is required to discriminate between 0 and $\pm 1$. However, it is also clear that the error is smaller for larger sizes of the modulation cell, indicating the presence of unwanted noise effects, like residual diffraction for instance. 

\subsection{Computation of the Normalized Trace of a Matrix}
As we can see from the results in Eq. (\ref{traco}), we can implement the DQC1 model
for a general function programmed in the SLM, where we can define squares or other geometries 
composed by an arbitrary number of pixels and modulate each one with some arbitrary
phase ranging from $0$ to 2$\pi$. The measurement of $\langle \sigma_x \rangle$
gives the real part of the sum over all phases, and $\langle \sigma_y \rangle$
gives the imaginary part. Therefore, the system is able to perform the calculation of the
trace of a normalized matrix imprinted in the SLM. In this case, it is not only a matter of determining if
the function is balanced or constant, but rather of obtaining a result that is as precise
as possible. 

\begin{figure}[tbp] 
\leftline{\includegraphics[width=3.8in,height=3.86in,keepaspectratio]{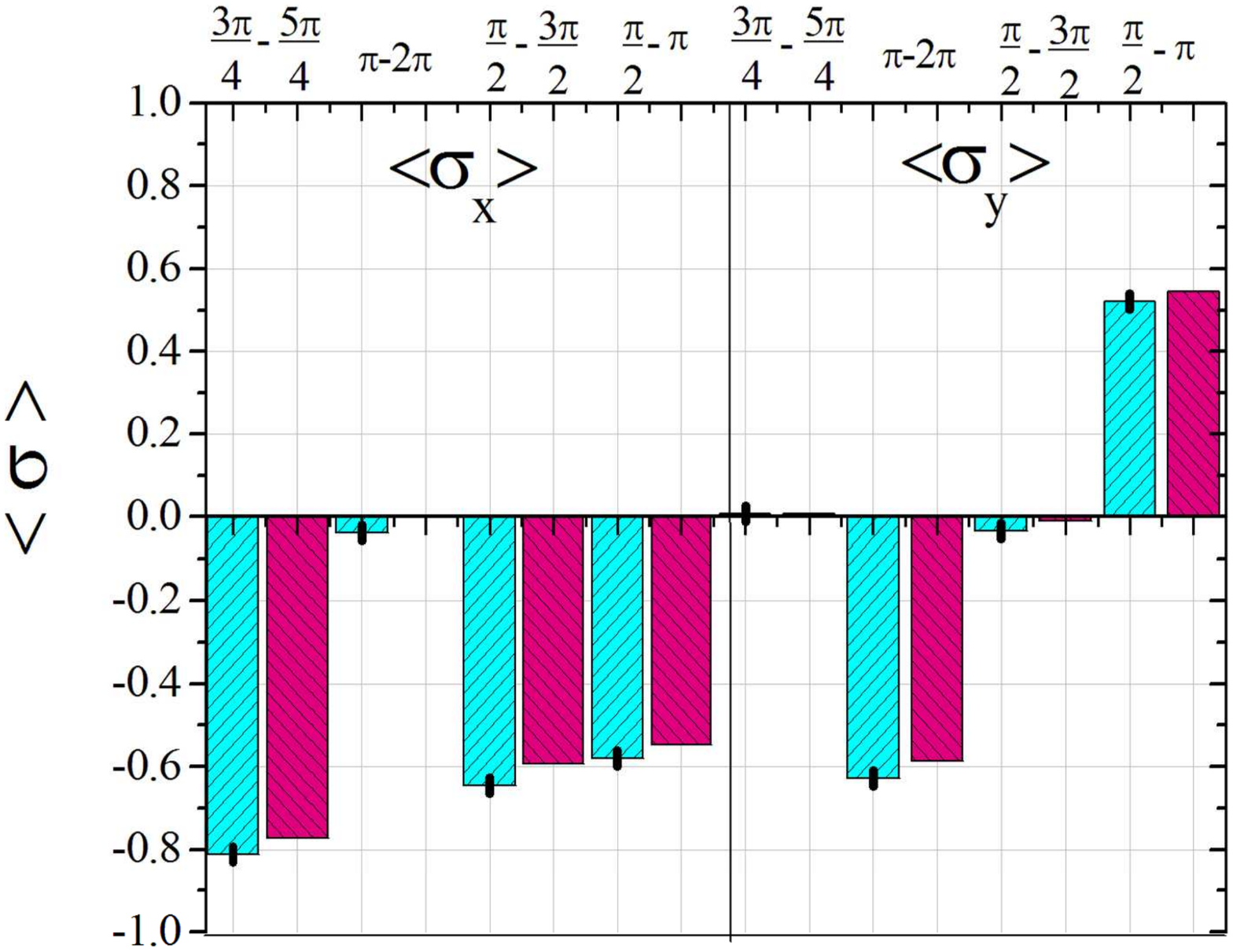}}
\vspace{-0.8cm}
  \caption{Experimental and theoretical results of the computation for the
traces of the matrices defined by Eq. (\ref{Eq:LinearPhases}). Columns filled 
with diagonal(/) lines (blue) represent experimental results of measurements of $\langle \sigma_x \rangle$ (left side) and $\langle \sigma_y \rangle$ (right side), and columns filled with anti-diagonal($\backslash$) lines (red) represent theory. The modulation ranges are displayed on the top.}
  \label{fig:Graph1}
\end{figure}

In order to illustrate the method, we programmed the SLM panel with phases varying linearly 
along the $y$ direction, according to:

\begin{equation}\label{Eq:LinearPhases}
\phi(x_i,y_j)=\phi_0 + \frac{y_j}{N_y}\phi_f,
\end{equation}
where $N_y=1080$ is the number of pixels along the $y$ direction. Thus, the matrix element corresponding to position $(x_i,y_j)$ is $e^{-i\phi(x_i,y_j)}$ . 
We have considered four cases of linear functions of the type given in Eq.~\ref{Eq:LinearPhases}, namely, with $(\phi_0,\phi_f)\;$=$\;(3\pi/4,5\pi/4)$, $(\pi,2\pi)$, $(\pi/2,3\pi/2)$, and $(\pi/2,\pi)$. The measured values $\langle\sigma_x\rangle_{exp}$ and $\langle\sigma_y\rangle_{exp}$ for the real and imaginary part of the trace of the corresponding matrices, respectively, are shown in Fig.~\ref{fig:Graph1} and summarized in Table I.

We compared the experimental values $\langle\sigma_x\rangle_{exp}$  and $\langle\sigma_y\rangle_{exp}$ with a theoretical prediction that takes into account the non-uniform intensity distribution of the light on the SLM panel combined with the unavoidable dephasing effect that is inherent to our Holoeye modulator, as described in detail in Ref.~\cite{105}. The theoretical prediction for the real and imaginary parts of the trace of the matrix is well described by

\begin{eqnarray}
\langle\sigma_x \rangle_{theo}=(1-2p)\sum_{i=1}^{N_x} \sum_{j=1}^{N_y} c_{ij} \cos[\phi(x_i,y_j)],\nonumber\\
\langle\sigma_y \rangle_{theo}=(1-2p)\sum_{i=1}^{N_x} \sum_{j=1}^{N_y} c_{ij} \sin[\phi(x_i,y_j)],
\label{theo}
\end{eqnarray}
where $p$ is a parameter which gives the degree of dephasing and $c_{ij}$ is the measured intensity of the light incident on the SLM pixel at position $(x_i,y_j)$. The overall dephasing effect is to decrease the coherences in Eq.~(\ref{rhopol}) by a factor of $(1-2p)$. According to Ref.~\cite{105}, the estimated value for $p$, considering the specific modulator we are using, is $0.08\pm0.02$. We determine $c_{ij}$ from the measurement of the transverse profile of the beam, shown in Fig.~\ref{beam}(b). The resolution of our measurement of the beam transverse intensity is set by the size of a square cell composed by $80\times 80$ pixels. Assuming a flat intensity distribution within the cell at position $(X_I,Y_J)$, 
the intensity $c_{ij}$ on a pixel at position $(x_i,y_j)$ inside the cell can be approximated by
\begin{equation}
\label{Eq:pij}
c_{ij}=\frac{1}{N}\frac{C_{IJ}}{80\times80},
\end{equation}
where $C_{IJ}$ is the coincidence counts for the corresponding cell and $N$ is a normalization factor given by
\begin{eqnarray}
N=\sum_{I=1}^{N_x/80}\sum_{J=1}^{N_y/80} C_{IJ}, 
\end{eqnarray}
so  that 
\begin{eqnarray}
\sum_{i=1}^{N_x} \sum_{j=1}^{N_y} c_{ij}=1.
\end{eqnarray}

For the sake of completeness, the purely theoretical values for the real and imaginary parts of the normalized trace of matrix M (${\rm Tr_{n}(M)}$), supposing the transverse field distribution to be flat in the entire SLM panel and disregarding the dephasing effect \cite{105}, given by

\begin{eqnarray}
{\rm Tr_{n}(M)} \equiv {\rm Tr(M)}/(N_x.N_y),
\label{puretheo}
\end{eqnarray}
are also shown in Table I.

\vspace*{.3cm}
\begin{table}[h!]
\begin{tabular}{|l|c|c|c|c|}
\hline
$(\phi_0,\phi_f)$ & $(\frac{3\pi}{4},\frac{5\pi}{4})$ & $(\pi,2\pi)$  & $(\frac{\pi}{2},\frac{3\pi}{2})$ & $(\frac{\pi}{2},\pi)$ \\ \hline
$\langle\sigma_x\rangle_{exp}$ & $-0.811$ & $-0.039$  & $-0.646$ & $-0.579$ \\ \hline
$\delta_{\langle\sigma_x\rangle}$ & $\pm 0.005$ & $\pm 0.008$  & $\pm 0.006$ & $\pm 0.007$ \\ \hline
$\langle\sigma_x\rangle_{theo}$ & $-0.773$ & $0.002$  & $-0.593$ & $-0.548$ \\ \hline
${\rm Re[Tr_{n}(M)]}$ & $-0.903$ & $-0.004$  & $-0.644$ & $-0.638$ \\ \hline

$\langle\sigma_y\rangle_{exp}$ & $0.007$ & $-0.628$  & $-0.034$ & $0.521$ \\ \hline
$\delta_{\langle\sigma_y\rangle}$ & $\pm 0.008$ & $\pm 0.007$  & $\pm 0.008$ & $\pm 0.007$ \\ \hline
$\langle\sigma_y\rangle_{theo}$ & $0.008$ & $-0.587$  & $-0.009$ & $0.545 $ \\ \hline
${\rm Im[Tr_{n}(M)]}$ & $0.012$ & $-0.637$  & $-0.003$ & $0.639$ \\ \hline
\end{tabular}
\label{Tabela1}
\caption{Experimental and theoretical results for the computation of the
traces of the matrices defined by Eq. (\ref{Eq:LinearPhases}).}
\end{table}

The error bars for the experimental values, shown in Table I, were computed considering the error in the phase modulation to be equal to the smallest 
modulation step, which is $\delta\phi \simeq 2\pi/256$, and considering the error in the intensity of
the light in the SLM given by the uncertainty of the photon number distribution, considered to be Poissonian,
so that $\delta c_{ij} \simeq \sqrt{c_{ij}}$. In general, we have a rather good agreement between experimental and theoretical values within the technical limitations. For instance, they are related to the intrinsic discretization of the modulator (here, we refer to the 256 levels of phase modulation), diffraction, inhomogeneities in the optical beam profile, and imperfections of the optical devices such as waveplates and polarizing beam splitters. Nevertheless, the results serve for a successful proof of principle.


\section{Conclusion}

In conclusion, we present an experiment in which the DQC1 quantum computation
model is implemented using a polarization-controlled spatial light modulator (SLM) 
acting on the wavefront of single-photon fields. 
We illustrate the utility of the system by implementing the Deutsch-Jozsa algorithm
on a system whose size is the equivalent of about 19 qubits using a matrix of 960 x 960. 
The resolution of the SLM allows for the representation of the equivalent of up to about 21 qubits.
We also show that a more general calculation is possible and experimentally compute, through polarization measurements, the normalized trace of a matrix whose diagonal elements are represented by modulation phases.

A future path to improve our results concerns the use of alternative optical devices
that could also implement polarization controlled operations on the transverse spatial 
degrees of freedom of photons with better performance. One possibility could be the improvement
of the SLM technology or the combined use of multiple SLMs. In this case we might be able to
handle much larger matrices. 

 \begin{acknowledgments}
 Financial support was provided by Brazilian agencies CNPq, 
CAPES, FAPERJ, and the National Institute of Science and Technology for Quantum Information.
 \end{acknowledgments}

\end{document}